\documentclass[superscriptaddress,showpacs,twocolumn]{revtex4}
\usepackage{amssymb}
\usepackage[tbtags]{amsmath}
\usepackage{graphicx}
\usepackage{epsfig,graphicx,times}

\begin{document}

\title{Joint quantum nondemolition measurements of qubits: beyond the mean-field
theory}
\author{L. F. Wei}
\affiliation{Quantum Optoelectronics Laboratory, School of Physics
and Technology, Southwest Jiaotong University, Chengdu 610031,
China}
\affiliation{Centre for Quantum Technologies and Physics
Department, National University of Singapore, 2 Science Drive 3,
Singapore 117542}
\affiliation{State Key Laboratory of
Optoelectronic Materials and Technologies, School of Physics and
Engineering, Sun Yat-Sen University, Guangzhou 510275, China}
\author{J. S. Huang}
\affiliation{Quantum Optoelectronics Laboratory, School of Physics
and Technology, Southwest Jiaotong University, Chengdu 610031,
China}
\author{X. L. Feng}
\affiliation{Centre for Quantum Technologies and Physics Department,
National University of Singapore, 2 Science Drive 3, Singapore
117542}
\author{Z. D. Wang}
\affiliation{Department of Physics and Center of Theoretical and
Computational Physics, The University of Hong Kong, Pokfulam Road,
Hong Kong, China}
\author{C. H. Oh}
\affiliation{Centre for Quantum Technologies and Physics Department,
National University of Singapore, 2 Science Drive 3, Singapore
117542}

\date{\today}

\begin{abstract}
We propose an approach to nondestructively detect $N$ qubits by
measuring the transmissions of a dispersively-coupled cavity. By
taking into account all the cavity-qubits quantum correlations
(i.e., beyond the usual coarse-grained/mean-field approximations),
it is revealed that for an unknown normalized $N$-qubit state
$|\psi_N\rangle=\sum_{k=0}^{2^N-1}\beta_k|k\rangle_N$, each detected
peak in the cavity transmitted spectra marks one of the basis states
$|k\rangle_N$ and the relative height of such a peak is related to
the corresponding superposed-probability $|\beta_k|^2$.
Our results are able to unambiguously account for the intriguing
multi-peak structures of the spectra observed in a very recent
circuit-quantum-electrodynamics experiment [Phys. Rev. A {\bf 81},
062325 (2010)] with two superconducting qubits.

\end{abstract}

 \maketitle 
\vspace{-0.2cm}

{\it Introduction.---} It is well-known that the readout of qubits
is one of necessary steps in quantum information processing.
Phenomenally, the information stored in an unknown N-qubit quantum
state $|\psi_N\rangle=\sum_{k=0}^{2^N-1}\beta_k|k\rangle_N$ can be
partly extracted by directly applying the standard von Neumann
projective operation $\hat{P}=\sum_k|k\rangle_{NN}\langle k|$ to the
quantum register~\cite{book}. After such a projection, the register
is collapsed to one of the computational basis (basis states)
$\{|k\rangle_N=|\sum_{j=1}^N2^{j-1}\alpha_j\rangle_N
=\prod_{j=1}^N|\alpha_j\rangle_N, \alpha=0,1\}$ with a probability
$|\beta'_k|^2$. This is a directly projective measurement (DPM) and
the register is detected. Typically, DPM has been utilized to
extract the binary quantum information stored in~\cite{ion-rmp},
such as trapped ions, Cooper-pair boxes, and the current-biased
Josephson junctions, etc..
Essentially, due to the inevitable back actions of the the measuring
apparatus, the detected $|\beta'_k|^2$ is always less than its
expectable value $|\beta_k|^2$. This means that the DPM is not an
ideal method to extract the quantum information in an unknown
quantum state~\cite{Ralph06}.

\begin{figure}[htbp]
\includegraphics[width=8cm,height=3cm]{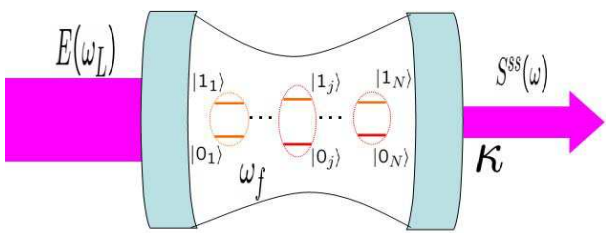}
\caption{(Color online) QND measurements of $N$
qubits (with two levels: $\{|0_j\rangle,|1_j\rangle\}, j=1,...,N$)
by measuring the steady-state transmitted spectra
$S^{ss}_N(\omega_L)$ of the dispersively-coupled cavity (with
frequency $\omega_f$ and decay rate $\kappa$) driven by a
frequency-controllable external field $E(\omega_L)$.}
\end{figure}

Alternatively, indirectly projective measurements (IPMs) can also be
utilized to achieve the measurement of the qubits, where another
coupled system instead the qubits-selves is detected. Typical
examples include, e.g., dc-SQUIDs for the inductively-connected
Josephson flux (persistent) qubits~\cite{mooij10}, optical cavities
for the containing atomic qubits~\cite{cQED}, and Cooper-pair box
for the nanomechanical resonators~\cite{wei-prl}, etc..
A remarkable advantage in the IPM is that the back actions from the
detected system could be minimized. If the condition $[H_N, H_I]=0$
is satisfied (which means that the disturbance of the detector $D$
on the qubits is negligible), then the relevant IPM further becomes
a quantum nondemolition (QND) measurement~\cite{book}. Here, $H_N$
is the Hamiltonian of the N-qubit register and $H_I$ the interaction
between it and the detector. Note that the term nondemolition does
not imply that the wave function of the register fails to collapse
due to the measurement~\cite{Wahyu}. In fact, for an unknown input
state $|\psi_N\rangle$, after the QND measurement the N-qubit
register will be automatically collapsed to one of its computational
basis $|k\rangle_N$ with an {\it ideal} probability $|\beta_k|^2$.
Thus, QND measurement is a conceptually ideal projective
measurement; the successive QND measurements on the same register
will give the same result.

As it can be easily detected with the current technique, driven
cavity has been widely utilized to achieve the desirable IPM of the
dispersively-coupling qubits. Experimentally, it is not difficult to
probe the resonance frequency of a driven cavity by detecting the
transmitted signals. If the qubits are dispersively coupled to the
cavity mode, then the cavity is pulled by the qubits, depending on
the state of the qubits. As a consequence, by detecting the shift of
the central frequency of the driven cavity mode the QND measurement
of the qubits can be achieved. This idea has been experimentally
demonstrated by the cavity QED experiments with few qubits, and
single basis states (i.e., computational basis) of the atomic and
superconducting qubits had been experimentally
distinguished~\cite{kimble, Wallraff}.
Next challenge is to completely characterize an unknown N-qubit
state $|\psi_N\rangle$ by nondestructively measuring an arbitrary
superposition of the single basis states with ideal precisions.

Considering the practically-existing dissipation of the detector
(i.e., cavity) and also the statistical quantum correlations between
it and the $N-$qubit quantum register, in this letter we show that
an unknown quantum state $|\psi_N\rangle$ can be effectively
nondestructively detected by the realistic QND measurements.
Our proposal still works for the mixed states, and thus could be
utilized to explain the detected multi-peak structure in the most
recent circuit-quantum-electrodynamics (circuit-QED)
experiment~\cite{Chow10}, where the detected qubits could be
prepared/decayed at various superpositions of all the possible basis
states.

{\it Generic model.---}The system proposed to nondestructively
detect an $N$-qubit state is schematized in Fig.~1, wherein $N$
non-interacting qubits are dispersively coupled to a driven cavity.
Certainly, the preparation of the initial state of the qubits and
detection of the driven cavity are repeatable.
Without loss of generality, we assume that, the coupling strength
$g_j$ and the detuning $\Delta_j=|\omega_f-\omega_j|$ between the
$j$th qubit and the cavity, and the detuning
$\Delta_{ij}=|\omega_i-\omega_j|$ between the $i$th and $j$th qubits
satisfy the condition
\begin{equation}
0<\frac{g_j}{\Delta_j},\,\frac{g_ig_j}{\Delta_i\Delta_{ij}},\,\frac{g_ig_j}{\Delta_j\Delta_{ij}}\ll
1,\,\,i\neq j=1,2,...,N.
\end{equation}
This is to realize the dispersive interactions between the qubits
and cavity, and to assure that the $i$th and $j$th qubits are
decoupled effectively from each other.
Also, the decay rates $\{\gamma_j\}$ of the qubits should be
significantly less than that of the cavity, $\kappa$, to ensure that
the detected state of the qubits has sufficiently-long lifetime. In
fact, all the conditions listed above are practically satisfied
 in the current typical circuit-QED systems~\cite{Wallraff}.

Under the rotating-wave approximation and in a framework rotating at
a frequency $\omega_L$, the process for nondestructively measuring
the unknown $N$-qubit state could be described by the following
master equation
\begin{eqnarray}
\dot{\rho}_N&=&-i[H_N,\rho_N]
+\frac{\kappa}{2}(2\hat{a}\rho_N\hat{a}^\dagger-\hat{a}^\dagger\hat{a}\rho_N
-\rho_N\hat{a}^\dagger\hat{a}),\\
H_N&=&\delta\hat{a}^\dagger\hat{a}
+\frac{1}{2}\sum_{j=1}^N\tilde{\omega}_j\sigma_j^z
-\hat{a}^\dagger\hat{a}\sum_{j=1}^N\Gamma_j\sigma_j^z
+\epsilon(\hat{a}^\dagger+\hat{a}),\nonumber
\end{eqnarray}
with $\Gamma_j=g_j^2/\Delta_j$ and $\epsilon$ being the effective
strength of the driving. Also, $\tilde{\omega}_j=\omega_j-\Gamma_j$
is the renormalized transition frequency of the $j$th qubit, and
$\delta=\omega_f-\omega_L$ is the detuning between the driving field
and the cavity.
For the ideal readouts, the dissipation of the qubits is assumed to
be negligible. As a consequence, the expectable values of all the
related qubit-operators, i.e., $\sigma^z_j$, $\sigma^z_j\sigma^z_k$
($j\neq k$), and the $N$-body ones $\prod_{j=1}^N\sigma^z_j$ are
kept unchanged. Our central task is to calculate the steady-state
transmitted strength $S_N^{ss}(\omega_L)$ of the driven cavity. This
quantity is essentially proportionate to the number of the
steady-state photons in the cavity, i.e.,
$S_N^{ss}=\langle\hat{a}^\dagger\hat{a}\rangle_N^{ss}/\epsilon^2$,
which is determined by the following dynamical equation
\begin{eqnarray}
\frac{d\langle\hat{a}^\dagger\hat{a}\rangle_N}{dt}
=-\kappa\,\langle\hat{a}^\dagger\hat{a}\rangle_N
-2\epsilon\mathrm{Im} \langle\hat{a}\rangle_N.
\end{eqnarray}
Here, $\langle\hat{a}\rangle_N={\rm Tr}(\hat{a}\rho_N)$ is further
determined by
\begin{eqnarray}
\frac{d\langle\hat{a}\rangle_N}{dt}=\left(-i\delta-\frac{\kappa}{2}\right)\langle\hat{a}\rangle_N
+i\sum_{j=1}^N\Gamma_j\langle\sigma_j^z\hat{a}\rangle_N -i\epsilon.
\end{eqnarray}

Neglecting all the statistical quantum correlations between the
cavity and qubits, i.e., under the usual coarse-grained (or
mean-field) approximation (CGA), see, e.g.,~\cite{Bianchetti}, we
simply have $\langle\sigma_j^z\hat{a}\rangle_N\approx
\langle\sigma_j^z(0)\rangle_N \langle\hat{a}\rangle_N$. Then, by
finding the steady-state solutions to the Eqs.~(3-4), one can easily
obtain an approximate transmitted spectrum:
\begin{equation}
\tilde{S}^{ss}_N(\omega_L)=
\left\{\left[\omega_L-\left(\omega_f-\Delta\tilde{\omega}_N\right)\right]^2
+\left(\frac{\kappa}{2}\right)^2\right\}^{-1},
\end{equation}
with $\Delta\tilde{\omega}_N=\sum_{j=1}^N
\Gamma_j\langle\sigma_j^z(0)\rangle_N$. This indicates that,
compared to the spectrum for the empty cavity (EMC) transmission,
the qubits only shift the central frequency with a quantity
$\Delta\tilde{\omega}_N$ and the single-peak shape is unchanged.
However, the above CGA is unnecessary and the two-body cavity-qubits
correlation functions $\langle\sigma_j^z\hat{a}\rangle_N$ can be
further determined by solving the following dynamical equation
\begin{equation}
\frac{d\langle\sigma_j^z\hat{a}\rangle_N}{dt}=\left(-i\delta-\frac{\kappa}{2}\right)
\langle\sigma_j^z\hat{a}\rangle_N
-i\epsilon\langle\sigma_j^z\rangle_N
+i\sum_{l=1}^N\Gamma_l\langle\sigma_l^z\sigma_j^z\hat{a}\rangle_N.
\end{equation}
Note that the three-body cavity-qubits correlations
$\langle\sigma_j^z\sigma_l^z\hat{a}\rangle_N$, introduced above, is
related further to the four-body cavity-qubits correlations:
$\langle\sigma_j^z\sigma_l^z\sigma_m^z\hat{a}\rangle_N$,
$m=1,2,...,N$, etc..
Generally, the $k$-body cavity-qubits correlations are related
further to the $(k+1)$-body cavity-qubits correlations (i.e., $k$
qubits correlate simultaneously to the cavity), and thus a series of
dynamical equations for these correlations will be induced.
Fortunately, due to the fact that $\sigma_l^z\sigma^z_m=1$ for
$l=m$, these equation-chains will be automatically cut off and ended
at the $(N+1)$-body cavity-qubits correlations. Then, all the
interested statistical quantum correlations in these equations can
be exactly calculated, and consequently the transmitted spectra can
be obtained beyond the usual CGAs.
It is emphasized that the spectral distribution $S_N^{ss}(\omega_L)$
including all the cavity-qubits quantum correlations may reveal
$2^N$ peaks for an $N$-qubit state superposed by $2^N$ basis states.
If the detected state is just one of the basis states (not their
superposition), then $S_N^{ss}(\omega_L)$ reduces to
$\tilde{S}_N^{ss}(\omega_L)$ (with a single-peak structure) and the
cavity-qubits quantum correlation vanishes.

{\it Demonstrations with experimentally-existing circuit-QED
systems.---}Our generic proposal derived above could be specifically
demonstrated with various experimental cavity-qubits systems,
typically the circuit-QED one~\cite{Wallraff,Filipp}. In this
system, the cavity is formed by a coplanar waveguide (of the length
at the order of millimeters) and the qubits are generated by the
Cooper-pair boxes (CPBs) with controllable Josephson energies. At a
sufficiently low temperature (e.g., $\leq20$ mK), the coplanar
waveguide works as an ideal superconducting transmission line
resonator (i.e., cavity). Experimentally~\cite{Bianchetti}, the
decay rate (e.g., $\kappa=2\pi\times 1.69$MHz) of the cavity is
about ten times larger than that of the CBP-qubit (e.g.,
$\gamma=2\pi\times 0.19$MHz)~\cite{note}. Also, by adjusting the
external biases, the CPB-qubits could be either coupled to or
decoupled from the resonator, and the required initial state
preparation and detection can be robustly repeated.
For the EMC case, the steady-state solutions to Eqs.~(3-4) can be
easily obtained and the transmission spectrum reads
$S_{0}^{ss}(\omega_L)=[(\omega_L-\omega_f)^2 +(\kappa/2)^2]^{-1}$.
%
Obviously, this is a well-known Lorentzian lineshape~\cite{Wallraff}
centered at $\omega_f$, with the half-width $\kappa$.

For one qubit case with $N=1$, the steady-state transmission
spectrum of the cavity is expressed as
\begin{equation}
S_{1}^{ss}(\omega_L)=\frac{(\omega_L-\omega_f)^2
-2(\omega_L-\omega_f)\Gamma_1Z_1^{(1)}(0)+\Lambda_1}
{\left[(\omega_L-\omega_f)^2-\Lambda_1^2\right]^2
+\left[\kappa(\omega_L-\omega_f)\right]^2},
\end{equation}
with $\Lambda_1=\Gamma_1^2+(\kappa/2)^2$ and $Z_1^{(1)}(0)={\rm
Tr}\{\rho_1(0)\sigma_1^z\}=2|\beta_1|^2-1$, for the unknown qubit
state $|\psi_1\rangle$. This is evidently different from the
$\tilde{S}_1^{ss}(\omega_L)$ derived under the usual CGA.
Obviously, the spectrum function $S_1^{ss}(\omega_L)$ predicates
that two transmitted peaks could be found in the spectrum.
Specifically, using the parameters in the experimental circuit-QED
with one CPB-qubit~\cite{Bianchetti}:
$(\omega_f,\omega_0,\kappa,g)=2\pi\times(6444.2,4009,1.69,134)$MHz,
we plot respectively the spectra $\tilde{S}_1^{ss}(\omega_L)$ and
$S_1^{ss}(\omega_L)$ in Figs.~2(a) and 2(b) typically for
$|\beta_1|^2=0, 0.25, 0.5, 0.75,$ and $1$, respectively. For
contrasts, the spectrum of the empty cavity (black line) is also
plotted in the figures. We make two remarks. Firstly, if the qubit
is at one of their basis states (i.e., either $|0\rangle_1$ or
$|1\rangle_1$), then $\tilde{S}_1^{ss}(\omega_L)$ and
$S_1^{ss}(\omega_L)$ give the same single-peak distribution, which
has been experimentally demonstrated~\cite{Wallraff}.
Secondly, if the qubit is prepared beforehand at the superposition
of its two basis states, then $\tilde{S}_1^{ss}(\omega_L)$ shows
still the single-peak structure (when $|\beta_1|=0.5$,
$\tilde{S}_1^{ss}(\omega_L)$ superposes the $S_0^{ss}(\omega_L)$)
but $S_1^{ss}(\omega_L)$ predicts two peaks: the locations of the
central frequencies are unchanged, but their relative heights equal
respectively to the superposed probabilities of the two basis
states. Therefore, $S_1^{ss}(\omega_L)$ (rather than
$\tilde{S}_1^{ss}(\omega_L)$) provides the messages of all the
diagonal elements of the density matrix $\rho_1$. These predictions
should be easily verified with the current experimental technique,
once the qubit is input at an arbitrarily-selected superposition
state.

\begin{figure}[htbp]
\vspace{0.25cm}
\includegraphics[width=8.1cm,height=3cm]{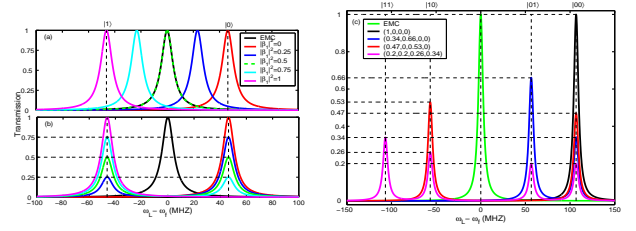}
\caption{(Color online) Left: Spectral distributions
$\tilde{S}_1^{ss}(\omega_L)$ (a) and $S_1^{ss}(\omega_L)$ (b) of the
cavity with a single qubit prepared at the state
$|\psi\rangle_1=\beta_0|0\rangle_1+\beta_1|1\rangle_1$, typically
for $|\beta_1|^2=0, 0.25, 0.5, 0.75$, and $1$, respectively. Right:
Spectral distributions $S_2^{ss}(\omega_L)$ for the two-qubit
register prepared at $|\psi_2\rangle$ with $(|\beta_0|^2,
|\beta_1|^2, |\beta_2|^2, |\beta_3|^2)=(1, 0, 0, 0), (0.34, 0.66, 0,
0), (0.47, 0, 0.53, 0), (0.2, 0.2, 0.26, 0.34)$, respectively. In
contrast, transmission spectrum of the empty cavity (EMC)
$S_0^{ss}(\omega_L)$ is also shown by the black line.}
\end{figure}

Similarly, for $N=2$ case~\cite{Chow10, nature09, Filipp} the
steady-state transmitted spectrum can still be analytically
obtained:
\begin{eqnarray}
S_2^{ss}(\omega_L)=-\frac{2(AC+BD)}{\kappa(A^2+B^2)},
\end{eqnarray}
with
$
A=(\Gamma_1^2-\Gamma_2^2)^2+2[\frac{\kappa^2}{4}-(\omega_L-\omega_f)^2]\sum_{j=1}^2\Gamma_j^2
+[\frac{\kappa^2}{4}-(\omega_L-\omega_f)^2]^2-\kappa^2(\omega_L-\omega_f)^2,
$
$B=-2\kappa(\omega_L-\omega_f)[\sum_{j=1}^2\Gamma_j^2+\frac{\kappa^2}{4}-(\omega_L-\omega_f)^2],
$
$C=\,\kappa Z_{1,2}^{(2)}(0)\Gamma_1\Gamma_2
-\kappa(\omega_L-\omega_f)
\sum_{j=1}^2\,\langle\sigma_j^z(0)\rangle_2\Gamma_j
+\frac{\kappa}{2}[3(\omega_L-\omega_f)^2-\frac{\kappa^2}{4}-\sum_{j=1}^2\Gamma_j^2],\,\,
Z_{1,2}^{(2)}(0)={\rm Tr}\{\rho_2(0)\sigma_1^z\sigma_2^z\},\,
Z_j^{(2)}(0)={\rm Tr}\{\rho_2(0)\sigma_j^z\}$
and $ D=-2\,Z_{1,2}^{(2)}(0)(\omega_L-\omega_f)\Gamma_1\Gamma_2
-\sum_{j=1}^2\,Z_j^{(2)}(0)\Gamma_j[\Gamma_j^2-\Gamma_{j'}^2
+\frac{\kappa^2}{4}-(\omega_L-\omega_f)^2]
+(\omega_L-\omega_f)[\sum_{j=1}^{2}\Gamma_j^2+\frac{3\kappa^2}{4}-(\omega_L-\omega_f)^2],\,j\neq
j'=1,2,$ respectively.
It is seen that, the spectral distribution $S_2^{ss}(\omega_L)$ may
reveal four peaks, but $\tilde{S}_2^{ss}(\omega_L)$ always shows
one-peak structure.
The current circuit-QED experiments with two
CPB-qubits~\cite{Chow10} could be utilized to verify the multi-peak
spectral distributions predicted above, once the CPB-qubits are
prepared at the superpositions of their basis states.
Specifically, with the experimental
parameters~\cite{Chow10,nature09}: $(\omega_f=\omega_c,
\Gamma_1=\chi^L, \Gamma_2=\chi^R,\kappa)=2\pi\times(6.806, 0.013,
0.004, 0.001)$GHZ, one can plot the spectral function
$S_2^{ss}(\omega_L)$ in Fig.~2(c) for the selected two-qubit state
$|\psi_2\rangle$ with: $(|\beta_0|^2, |\beta_1|^2, |\beta_2|^2,
|\beta_3|^2)=(1, 0, 0, 0)$ (black line), $(0.34, 0.66, 0, 0)$ (blue
line), $(0.47, 0, 0.53, 0)$ (red line), and $(0.2, 0.2, 0.26, 0.34)$
(pink line), respectively.
From these numerical results one can see that, if the two-qubit is
prepared at one of the basis states, e.g., $|00\rangle$ here, then
the transmitted spectrum of the driven cavity shows a single peak.
While, if the two-qubit are prepared at the any superposition of
their basis states, then the detected spectra should reveal
multi-peak structures, i.e., two peaks for black and red lines and
four peaks for the pink line.

Generally, if the N-qubit register is prepared at the superposition
of $M (\leq 2^N-1)$ basis states, then the cavity could be pulled by
$M$ forms and thus there are $M$ possible shifts of the cavity
resonance frequency. As a consequence, the detected cavity
transmitted spectrum will reveal $M$ peaks; the superposed
probability of one of the basis state determines the weight for
pulling the cavity and thus the relative height of the corresponding
transmitted peak.
Therefore, one could presume that the QND measurements of an
arbitrary $N$-qubit state could be achieved by analyzing the
transmission spectra of the dispersively-coupled cavity: from the
locations of the central frequencies of the detected peaks, one can
determine which basis states $\{|k\rangle_N\}$ are superposed; and
from the relative heights of the corresponding peaks, one can
determine the superposed probabilities $\{|\beta_k|^2\}$.

{\it Discussions and Conclusions.--} Our proposal is based on an
important assumption, i.e., each detection should be finished
sufficiently-fast such that the influence of the decay of the
detected quantum state is negligible. This condition is satisfied in
the current circuit-QED
experiment~\cite{Wallraff,Bianchetti,Chow10}, wherein each data for
recording the transmission event of light through the cavity can be
obtained in about 40 $ns$ and the decay time of the detected
CPB-qubit is, e.g., $T_1\sim 1\mu s$. Thus, the predicted multi-peak
transmitted spectra could be verified, once the CPB-qubits are
prepared at the superposition of various possible basis states (even
the mixed ones).

\begin{figure}[htbp]
\includegraphics[width=8.2cm,height=3.0cm]{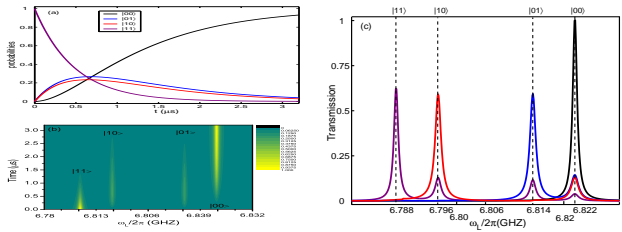}
\caption{(Color online) Left: (a) Occupation probabilities of
various basis states versus the decay time for the initial state
$|11\rangle$, (b) Spectral distributions versus the time for the
2-qubit register prepared at the basis state $|11\rangle$. Right:
(c) Time-averaged transmitted spectra over the time $\tau=0.5\,\mu$s
for different initial states: $|00\rangle$ (black), $|01\rangle$
(blue), $|10\rangle$ (red), and $|11\rangle$ (pink). Here, the decay
rate of the CPB-qubit is taken as $1/T_{1,j}\sim 1\,$MHz.}
\end{figure}

Immediately, our proposal could be utilized to explain the
multi-peak spectra observed in the recent experiment~\cite{Chow10}.
Strictly speaking, the time-averaged (over a relatively-long time,
i.e., $0.5\,\mu$s, the half of the decay time of the qubit) spectral
distributions shown there are not the desirable readouts of the two
CPB-qubits, due to the significant decays of the detected states
during these relatively-long measuring times for averages.
In Fig.~3(a) we typically show how the probabilities of various
basis states change with the time from the decays of the excited
state $|11\rangle$~\cite{note,Bianchetti}. Clearly, during the decay
(e.g., at $0.5\,\mu$s) the superpositions of the basis states are
induced. Thus, based on our proposal beyond the mean-field
approximation, the transmitted spectra would reveal a relevant
multi-peak structure. Phenomenally, the decay-time dependent
steady-state spectrum $S_2^{ss}(\omega_L,\tau)$ could be obtained by
replacing the unchanged expectable values of the qubit-operators in
Eq.~(8) (i.e., $Z_j^{(2)}(0)$ and $Z_{1,2}^{(2)}(0)$) with the
decay-time dependent ones~\cite{note,Bianchetti}.
Consequently, the time-dependent spectral distributions (due to the
decay of the initial state $|11\rangle$) was simulated in Fig.~3(b),
which agrees basically with the corresponding experimental
observation (i.e., Fig.~1(E) in Ref.~\cite{Chow10}).
Thus, except for the non-decayed ground state $|00\rangle$ (which
corresponds certainly to a single transmitted peak), the
time-evolution spectra due to the decay of arbitrary excited state,
e.g., $|10\rangle$\,\,(or $|01\rangle$, $|11\rangle$) would reveal
two (or two, four) peaks.
By integrating the decay-time dependent steady-state spectrum
$S_2^{ss}(\omega_L,\tau)$, Fig.~3(b) shows the relevant
time-averaged spectra (over the time interval $\tau=[0, 0.5]\,\mu$s)
for these time-evolutions.
One can see that the locations of the averaged peaks agree well with
the experimental observations~\cite{Chow10}. While, the relative
heights of the peaks (marking the basis states induced from the
decays of the input states) are relatively low. This is an
inevitable deduction of multiple QND measurements performed
sequentially within the averaged time.

In summary, an efficient approach to implement the QND joint
measurements of the $N$ qubits are proposed by detecting the
transmitted spectra through the dispersively-coupled cavity. These
measurements are the IPMs of the qubits, and thus the relevant
fidelities could be sufficiently high.
Remarkably, our proposal is a theory beyond the usual mean-field
approximation and thus the statistical quantum correlations between
the cavity and qubits are important. In deed, by specifically
solving the dynamical equation of the cavity-qubit correlations,
e.g., for $N=1$ case, one can prove that the lifetime of the
qubit-cavity quantum correlation is at the same order of the
cavity-self. Therefore, the effects of the cavity-qubit
correlations, i.e., the transmitted spectra with multiple peaks,
could be verified by inputting the superposed states of the qubits.

\vspace{-7mm}

\section*{Acknowledgments}
This work was supported in part by the National Science Foundation
grant No. 10874142, 90921010, and the National Fundamental Research
Program of China through Grant No. 2010CB923104, and A*STAR of
Singapore under research grant No. WBS: R-144-000-189-305. One of us
(Wei) thanks also Drs. Y. Yu and M.W. Wu for useful comments.


\begin{references}
\bibitem{note}Taking into account the dissipation $\{\gamma_j\}$ and
and dephasing $\{\gamma_{\phi,j}\}$ of the qubits, the proposed
spectral QND measurements should be described by a more generic
master equation:
%
$\dot{\rho}_N=-i[H_N,\rho_N]+\kappa \mathcal
{D}[a]\rho_N+\sum_{j}\gamma_{j} \mathcal {D}[\sigma_{-_j}]\rho_N
\nonumber+\sum_{j}\gamma_{\phi,j}\mathcal{D}[\sigma_{z_j}]\rho_N/2,
$
%
with the dissipation superoperator $\mathcal
{D}[A]{\varrho_N}=A{\varrho_N}A^\dagger-A^\dagger
A{\varrho_N}/2-{\varrho_N}A^\dagger A/2$. Specifically, the
expectable values of the qubit-operators are no longer unchanged but
determined by the equations (for a two-qubit system):
$d\langle\sigma_i^z\rangle/d\tau=-\gamma_{1,i}(\langle\sigma_i^z\rangle+1),\,i=1,2$,
and
$d\langle\sigma_1^z\sigma_2^z\rangle/d\tau=-(\sum_{i=1,2}\gamma_{1,i})
\langle\sigma_1^z\sigma_2^z\rangle-\sum_{i=1,2}\gamma_{1,i}\langle\sigma_i^z\rangle$.
%
For a single detection with the duration $T_e\sim 40$ns and the
qubit decay rate $\gamma_{1,j}^{-1}\sim 1\,\mu$s, the decay effects
is really negligible, i.e., the decay factor
$\exp(-\gamma_{1,j}T_e)\sim 1$. While, if the measurements went on a
relatively-long time, e.g., $\tau\sim 0.5\,\mu$s in
Ref.~\cite{Chow10}, to perform several QND-detections for time
averages, then the decay effects should be taken into account.
\end{references}
\end{document}